\documentclass{pasj00}

\begin{document}
\SetRunningHead{Y. Kato}{2:3 twin QPOs in MHD Accretion Flows}
\Received{2004/06/18}
\Accepted{2004/07/28}

\title{2:3 Twin Quasi-Periodic Oscillations in Magnetohydrodynamic
Accretion  Flows}

\author{Yoshiaki \textsc{Kato}
\thanks{Department of Physics, Graduate School of Science and
Technology, Chiba University, Inage-ku, Chiba 263-8522, Japan}}
\affil{Department of Physics, Graduate School of Science and
Technology, Chiba University, Inage-ku, Chiba 263-8522, Japan}
\email{kato@astro.s.chiba-u.ac.jp}

%


\maketitle

\begin{abstract}
We study the radial and vertical oscillations in the
three-dimensional magnetohydrodynamic (MHD) accretion flows around
black holes.  General relativistic effects are taken into account by
using pseudo-Newtonian potential.  We find that the structure of MHD
flows is changed at $3.8~r_{\rm s}\leq r\leq 6.3~r_{\rm s}$ and the
two pairs of quasi-periodic oscillations (twin QPOs) are excited in
that region.  The time evolution of the power spectrum density (PSD)
indicates that these twin QPOs are most likely to be produced by the
resonance between the Keplerian frequency $\nu_{\rm K}$ and the
epicyclic frequency $\nu_{\kappa}$.  The PSD shows that the lower peak
frequency $\nu_{\rm l}$ corresponds to $\nu_{\rm K}$, while the upper
peak frequency $\nu_{\rm u}$ corresponds to $\nu_{\rm K} +
\nu_{\kappa}$.  The ratio of two peak frequencies is close to 2:3.
The results provide the first direct evidence for the excitation of
the resonant disk oscillation in the MHD accretion flows.
\end{abstract}

\section{Introduction}

Quasi-periodic X-ray brightness oscillations (so-called QPOs) have
been observed in some low-mass X-ray binaries (LMXBs) including
neutron stars (NSs) and black hole candidates (BHCs).  After the
discovery of millisecond QPOs (so-called kHz QPOs) in NSs, the study
of kHz QPOs have been paid more attention.  Since their peak
frequencies are close to the orbital frequency at the innermost
circular orbit (ISCO) around NSs, the structures of both the disk and
the magnetosphere close to NSs can be obtained by studying their
origin (see van der Klis 1999 for a review).

In almost all kHz QPOs, the power spectrum density (PSD) shows twin
peaks whose frequencies are correlated with their X-ray intensity and
their peak separation are almost constant (e.g., Strohmayer et
al. 1996; Wijnands et al. 1997; Ford et al. 1997; Zhang et al. 1998).
In some NSs, a frequency separation of the twin peaks is comparable to
a frequency of a third QPO, which is detected during type I X-ray
bursts.

Since the third QPO is inferred to as a spin frequency of NSs, the
excitation of twin QPOs are conventionally explained by the
beat-frequency modulation between the Keplerian frequency of the disks
and the spin frequency of the NSs (so-called beat-frequency model: see
van der Klis 2000 for a review).  However, in contrast to the
beat-frequency model, the peak separations of some kHz QPOs are not
constant (e.g. Sco X-1: van der Klis 1997, 4U 1608-52: M\'{e}ndez et
al. 1998).

QPOs are also found in several BHCs (e.g., for GRS 1915+105: Morgan,
Remillard, and Greiner 1997, for XTE J1550-564: Remilard et al. 1999a,
for GRO J1655-40: Remillard et al. 1999b).  The PSD of QPOs in BHCs
shows single peak and its frequency is relatively stable.  In this
regards, QPOs in NSs and BHCs have been thought to have different
origins.  However, the recent discoveries of twin peak QPOs in BHCs
(e.g. Remillard et al. 2002) have cast some doubt on the distinction
of QPOs in NSs and BHCs.  Since the ratio of two peak frequencies is
about 2:3 in both BHCs and NSs, the origin of twin QPOs may be
identical to the accretion disk.

Interpretations of QPOs in the context of disk oscillations have been
developed by many authors (e.g., trapped disk oscillations: Kato~\&
Fukue 1980; Okazaki, Kato,~\& Fukue 1987; Nowak~\& Wagoner 1991, 1992,
resonances of disk oscillations: Abramowicz~\& Kluz\'{n}iak 2001; Kato
2003; Abramowicz et al. 2003).  These works investigate a growth of
unstable oscillation modes in geometrically thin disks around BHs in
order to produce a large fluctuations of X-ray intensity (see Kato
2001 for a review).

The first hydrodynamic simulations in the context of disk oscillations
were carried out by Matsumoto et al. (1988, 1989; see Kato, Fukue,~\&
Mineshige 1998 for a review).  They showed that acoustic oscillations
are trapped in the innermost region of the relativistic disk, when a
viscosity parameter $\alpha$ (Shakura~\& Sunyaev 1973) is large,
typically $\alpha\geq 0.1$.  As a result, a quasi-periodic oscillation
with the maximum of epicyclic frequency is established in that region.

Later, many hydrodynamic simulations were published (e.g., for
non-isothermal disks: Honma, Matsumoto,~\& Kato 1992; Chen~\& Taam
1995, for two-dimensional non-adiabatic disks: Milsom~\& Taam 1996,
1997).  Although these studies successfully demonstrated the
excitation of a trapped oscillation in the disk, the resonant
oscillations have a bit advantage to explain 2:3 twin QPOs.

In the framework of the resonant oscillations in relativistic disks,
Abramowicz et al. (2003) investigated the geodesic motion of particles
around BHs.  Later, Kato (2003) extended their idea into the disk
fluid.  They concluded that twin peak QPOs are excited by the
resonance of the disk oscillations.  They also asserted that the
resonant disk oscillation model can determine the mass and the spin
parameter of BHs by using the peak frequencies of twin QPOs (see also
Abramowicz~\& Kluz\'{n}iak 2001).

The study of the resonant oscillations in the disk have many
precedents.  Periodic brightness variations in dwarf nova were
investigated in the context of the tidal instability as a result of
the resonance in the disk (e.g., Whitehurst 1988; Hirose~\&Osaki 1990;
Lubow 1991a).  Hirose, Osaki, and Mineshige (1991) demonstrated the
tidal deformation of the disk as a result of the resonance in the disk
by using the three-dimensional smoothed particle hydrodynamics (SPH)
simulations (see also Lubow 1991b).

The previous simulations of disk oscillation assumed the
$\alpha$-viscosity model.  However, the MHD turbulence is now widely
accepted to be the source of disk viscosity since the Maxwell stress
coupled with the magnetorotational instability  (MRI: Balbus~\& Hawley
1991) can effectively transport angular momentum of the disk (e.g.,
Stone~\& Pringle 2001; Hawley, Balbus,~\& Stone 2001; Sano et
al. 2004).  Therefore the dynamics of the accretion flow is controlled
by MHD turbulence and the $\alpha$-viscosity calculations may fail to
represent the numerous aspects of the disk oscillations (see Ogilvie
2003).

Although many MHD disk simulations have been carried out so far
(e.g., Hawley 2000; Machida, Hayashi,~\& Matsumoto 2000; Hawley~\&
Krolik 2001; Hawley, Balbus,~\& Stone 2001; Armitage~\& Reynolds 2003;
Machida~\& Matsumoto 2003), the numerical study of MHD disk
oscillations has not been worked out yet.

In the present study, we perform three-dimensional MHD simulations of
a weakly magnetized rotating torus with the same initial condition as
that of Kato, Mineshige,~\& Shibata (2004, hereafter referred to as
KMS04).  The purpose of the present study is to examine the disk
oscillations in MHD accretion flows plunging into BHs.  To elucidate
MHD disk oscillations, we focus on the time evolution of the radial
structure of mass fluxes near the equatorial plane.  Our procedure
has an advantage to determine where QPOs are excited and what
frequency they have all together.  In \S 2, we describe methods of our
study.  We then present results in \S 3.  The final section is devoted
to discussion and summary.

\section{Methods}
We solve the basic equations of the resistive magnetohydrodynamics
in the cylindrical coordinates, $(r,\phi,z)$.  General relativistic
effects are incorporated by the pseudo-Newtonian potential
(Paczy\'{n}sky \& Wiita 1980), $\psi=-GM/(R-r_{\rm s})$, where
$R$($\equiv \sqrt{r^{2}+z^{2}}$) is the distance from the origin, and
$r_{\rm s}$ $(\equiv 2GM/c^{2})$ is the Schwarzschild radius (with $M$
and $c$ being the mass of a BH and the speed of light, respectively).
The basic equations and the physical conditions in the present study
are described in KMS04.

Hereafter we normalize all the lengths, velocities, and density by the
Schwarzschild radius, $r_{s}$, the speed of light, $c$, and the
maximum density of the initial torus, $\rho_{0}$, respectively (see
also KMS04).  The unit of the time is
$r_{s}/c=10^{-4}(M/10M_{\odot})\hbox{[sec]}$ when the mass of a BH is
$10M_{\odot}$.  For reader's convenience, the frequency is presented in
the unit of $10M_{\odot}/M~\hbox{[Hz]}$.

We start calculations with the same initial condition as that of model
B in KMS04.  Since model B does not present strong outflows, this model
is appropriate to evaluate how disk oscillations are excited in the
MHD accretion flows.  Taking into consideration of the flows crossing
the equatorial plane, we take away a symmetric boundary condition on
the equatorial plane in the previous calculation.  We impose the
absorbing inner boundary condition at the sphere
$R=\sqrt{r^{2}+z^{2}}=2$ (see KMS04 for more details).  In the present
calculations, we use $300\times 32\times 400$ non-uniform mesh points.
The grid spacing is the same as KMS04.  The entire computational box
is $0\leq r\leq 200$, $0\leq\phi\leq 2\pi$, and $-50\leq z\leq 50$ (in
comparison with KMS04, who solved only $0\leq z\leq 100$).

In order to examine disk oscillations, we focus on the structure of
the flow near the equatorial plane.  We calculate both radial and
vertical mass fluxes as follows:
\begin{equation}
\dot{m}_{\rm
r}(r,t)=-\int_{-5}^{5}dz\int_{0}^{2\pi}d\phi\int_{r}^{r+\Delta r}dr
[\rho v_{\rm r}],
\end{equation}
\begin{equation}
\dot{m}_{\rm
z}(r,t)=\int_{-5}^{5}dz\int_{0}^{2\pi}d\phi\int_{r}^{r+\Delta r}dr
[\rho v_{\rm z}].
\end{equation}
where $\rho$, $v_{\rm r}$ and $v_{\rm z}$ is the density, the radial
and vertical velocity, respectively.  Note that $\Delta r$ is the grid
spacing in the radial direction.

To quantify the oscillations of mass fluxes at different radii, we
employ the following quantities which are related to the gravitational
energy dissipated by the change of mass fluxes as functions of radius
and time,
\begin{equation}
L_{\rm r}(r,t)=\dot{m}_{\rm r}(r,t)\delta\psi(r),
\end{equation}
\begin{equation}
L_{\rm z}(r,t)=\dot{m}_{\rm z}(r,t)\delta\psi(r),
\end{equation}
where $\delta\psi(r)=d\psi(r)/dr$ is the gradient of the gravitational
potential $\psi(r)$ in the radial direction.

In the present study, we carry out the calculations until $\sim
3.2\times 10^{4}~\hbox{[$r_{s}/c$]} = 3.2~\hbox{[$(M/10M_{\odot})$
sec]}$ which correspond to more than $600$ orbital time at the ISCO
around $10M_{\odot}$ BH.  We sample the mass fluxes in the following
time interval:
\begin{equation}
\Delta t = 0.1\hbox{[$r_{\rm s}/c$]}={10\over
(10M_{\odot}/M)}\hbox{[$\mu$sec]}.
\end{equation}
This sampling rate is sufficient for the purpose of our study.

\section{Results}

\begin{figure*}[ht]
\begin{center}
\FigureFile(130mm,260mm){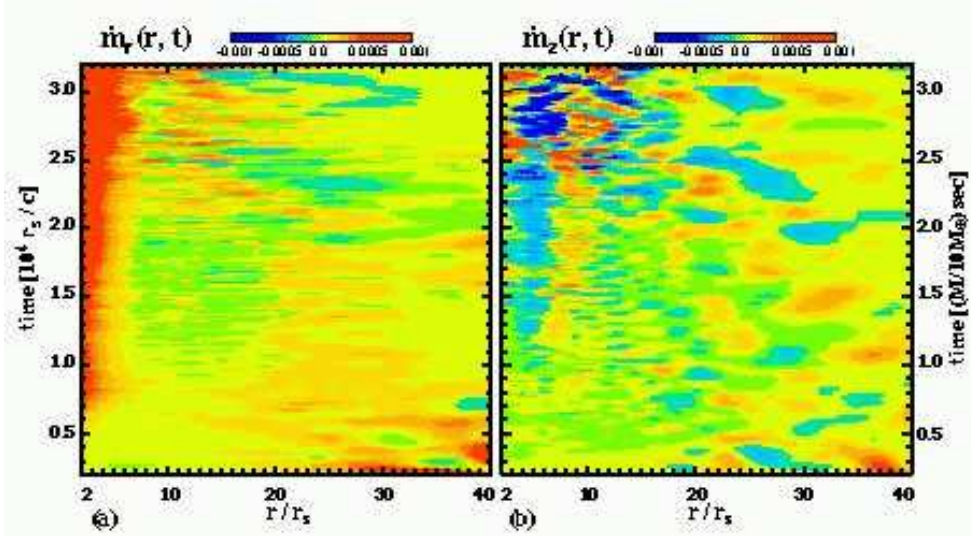}
\end{center}
\caption{The space-time diagrams of MHD accretion flows. (a) The
time evolution of the radial mass-flux.  The positive flux indicates
the inflow.  (b) The time evolution of the vertical mass-flux.  The
positive flux indicates the upward flow.  The flow patterns are
changed at  $r\sim 4 - 6$.  The fluctuations seems to be produced at
this region and they propagate outwards.}
\label{fig1:eps}
\end{figure*}

We first display the overall evolution of MHD accretion flows in
figure~\ref{fig1:eps}, in which the space-time diagrams of both radial
and vertical mass fluxes are shown in (a) and (b), respectively.  Mass
accretion takes place from the initial torus by the Maxwell-stress as
a result of the growth of MRI inside the torus.  After $t\sim
0.8$, the radial structure of the flow become quasi-steady and is
consistent with a hot, thick, subthermal, and near Keplerian disk
(KMS04; see also Hawley~\& Balbus 2002; Igumenshchev, Narayan,~\&
Abramowicz 2003).

The initial torus has totally been destroyed by the MHD turbulence
until $t\sim 1$.  Subsequently, the inner torus is created around
$r=25$ and it oscillates quasi-periodically (as is indicated by orange
color in the left panel).  The amplitude of both radial and vertical
mass fluxes seems to be increased after $t\sim 2$.  We can divide the
entire evolution of the MHD flows in the following two part: early
phase $t=1 - 2$ and late phase $t=2 - 3$.

The most interesting feature in the present simulations is that the
structures of the mass fluxes are changed at $r\sim 4 - 6$ in both
early and late phase.  In addition, the oscillations seems to be
excited in that region and they propagate outwards.

We compute the PSD of such variabilities of mass fluxes at all radii
by taking the Fourier transform of $L_{r}$ and $L_{z}$ as,
\begin{equation}
a(r,\nu)\approx\int_{t_{0}}^{t_{1}}L(r,t)\sin{[2\pi\nu(t-t_{0})]}dt,
\end{equation}
\begin{equation}
b(r,\nu)\approx\int_{t_{0}}^{t_{1}}L(r,t)\cos{[2\pi\nu(t-t_{0})]}dt,
\end{equation}
\begin{equation}
P(r,\nu)=\sqrt{[a(r,\nu)]^{2}+[b(r,\nu)]^{2}}.
\end{equation}
where $t_{0}$ ($t_{1}$) are the start (end) time of each phase, $\nu$
is the frequency ranging from $1$ to $10^{4}$ [Hz].  $L(r,t)$
represents the quantity of mass fluxes in either direction.
$P(r,\nu)$ is the PSD as functions of the radius and the frequency.

\begin{figure*}[ht]
\begin{center}
\FigureFile(130mm,260mm){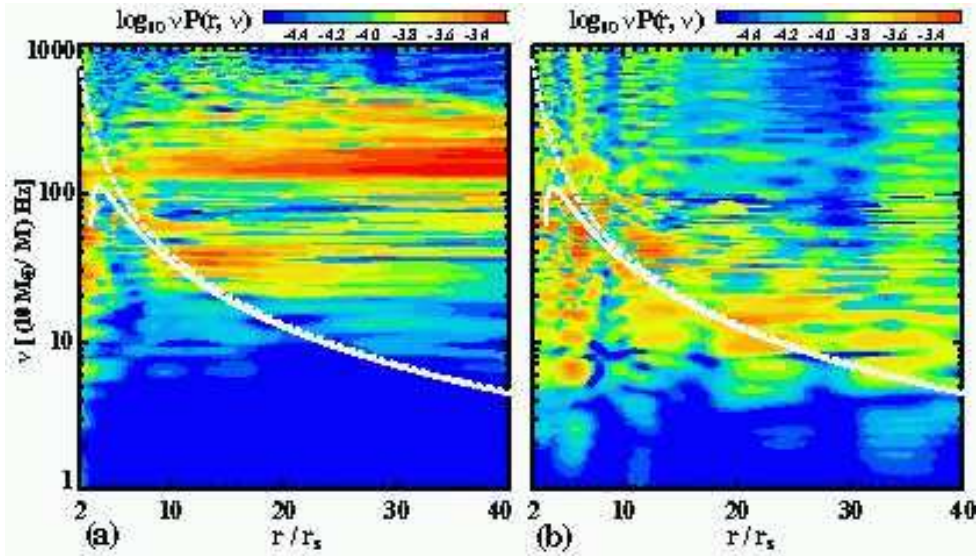}
\end{center}
\caption{The power spectral density (PSD) of (a) the radial
oscillation, (b) the vertical oscillation in the early phase.  The
color-contour indicates the distribution of PSD in logarithmic scale.
A solid curve indicates the epicyclic frequency $\nu_{\kappa}$ and a
dashed curve indicates the Keplerian frequency $\nu_{\rm K}$.  The
radial oscillation, whose frequency is about $150$, is distributed in
the wide range of the disk.  The vertical oscillations seems to be
distributed at the series of radii (e.g. $r\approx
6,\;13,\;20,\;35$).  These patterns are most likely to be produced by
the resonant disk oscillations.}
\label{fig2:eps}
\end{figure*}

Figure~\ref{fig2:eps} shows the contour of PSD of the radial and
vertical component in (a) and (b), respectively.  The color indicates
$\nu P(r,\nu)$ in logarithmic scale.  A dashed curve indicates the
Keplerian frequency $\nu_{\rm K}=\Omega/2\pi$ where $\Omega\equiv
1/\sqrt{2r(r-r_{s})^{2}}$.  A solid curve indicates the epicyclic
frequency $\nu_{\kappa}=\kappa/2\pi$ where
$\kappa\equiv\sqrt{2\Omega[2\Omega + r(d\Omega/dr)]}$.  Note that
these expressions are in the pseudo-Newtonian potential.  This figure
is useful to check a propagation region of radial and vertical
oscillations in the MHD flows.

In figure~\ref{fig2:eps}~(a), the most of PSD locates in the region
$\nu\geq \nu_{\kappa}$ where axisymmetric acoustic waves can
propagate.  The radial oscillation whose frequency about $150$ is
distributed in the wide range of the disk.  In addition, the radial
oscillations with broad range of frequency seems to be distributed in
the innermost region of the disk between the radius of the maximum
epicyclic frequency and the ISCO.  This component indicates that the
radial oscillations are trapped in that region.

Interestingly, the strong vertical oscillations with broad frequency
$5\leq\nu\leq 200$ are distributed at $r\sim 6$ in
figure~\ref{fig2:eps} (b).  Moreover, the vertical oscillations seems
to be distributed in the series of radii (e.g. $r\sim 6,~13,~22,~35$)
indicating that a pattern of PSD may be produced by the resonance in
the disk.

\begin{figure*}[ht]
\begin{center}
\FigureFile(130mm,260mm){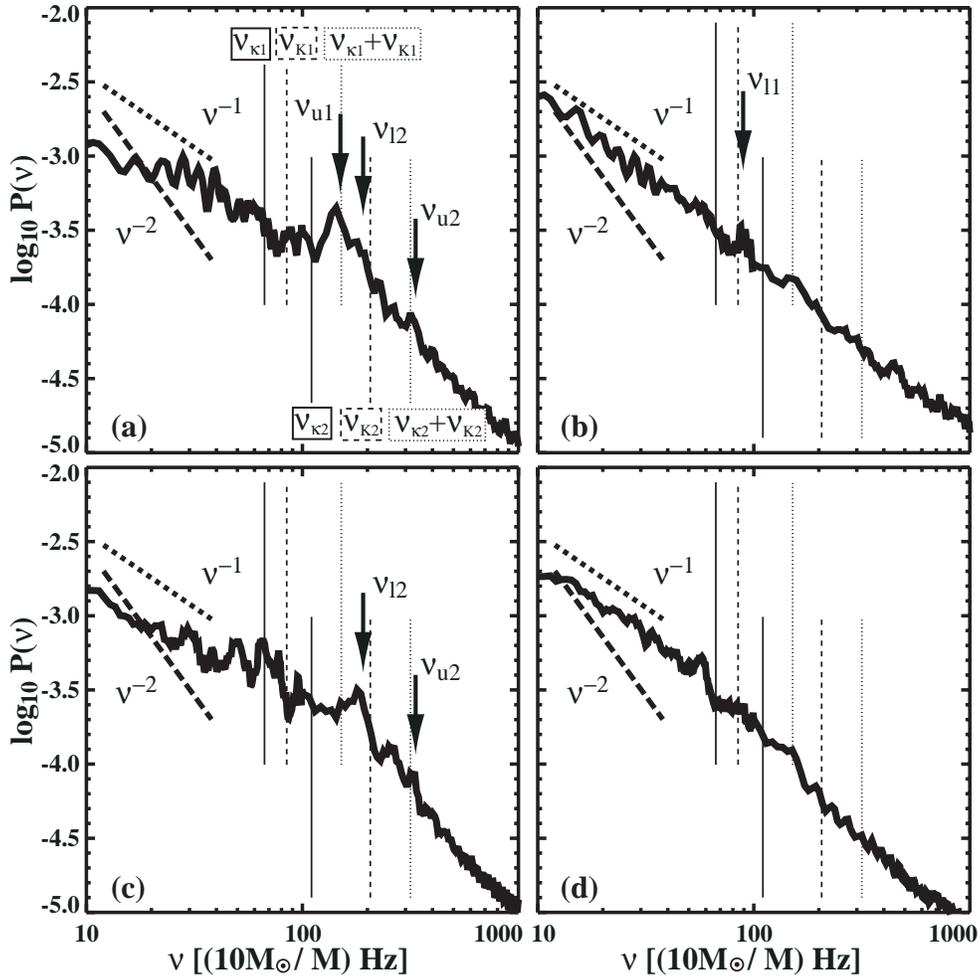}
\end{center}
\caption{The integrated PSD of (a) the radial oscillation, (b)
the vertical oscillation in early phase, whereas (c) the radial
oscillation, (d) the vertical oscillation in late phase.  A thick
dashed slope and a thick dotted slope indicate the frequency dependence
of $\nu^{-2}$ and $\nu^{-1}$, respectively.  Thin solid, dashed, and
dotted lines indicate the the epicyclic frequency $\nu_{\kappa}$, the
Keplerian frequency $\nu_{\rm K}$, the sum of them
$\nu_{\kappa}+\nu_{\rm K}$, respectively.  Their higher (lower)
frequencies are  at $r=3.8$ ($r=6.3$).  The QPOs are indicated by the
arrows.}
\label{fig3:eps}
\end{figure*}

To simulate the observed PSD of X-ray brightness variation, we
integrate $P(r,\nu)$ among the entire disk (from $r_{\rm in}=2$ to
$r_{\rm out}=200$), namely,
\begin{equation}
P(\nu)=\int_{r_{\rm in}}^{r_{\rm out}}P(r,\nu)dr.
\end{equation}
where $P(\nu)$ is the integrated PSD of the radial (vertical)
oscillations in the early phase is shown in figure~\ref{fig3:eps}~(a)
[(b)], whereas that in the late phase is shown in (c) [(d)].

Roughly speaking, all of the integrated PSD show power-law relations,
i.e., (a) $P(\nu)\propto \nu^{0}$ for $\nu < 40$, $\propto \nu^{-1}$
for $40\leq\nu < 200$, $\propto \nu^{-2}$ for $\nu\geq 200$, (b) and
(d) $P(\nu)\propto \nu^{-1}$ for all $\nu$, (c) $P(\nu)\propto
\nu^{-1}$ for $\nu < 200$, $\propto \nu^{-2}$ for $\nu\geq 200$.
These features are similar to that in the notion of self-organized
criticality (SOC: e.g., Mineshige, Takeuchi, and Nishimori 1994;
Takeuchi, Mineshige, and Negoro 1995).

We find a pair of {\it transient} QPOs (as indicated by the arrows
labeled as $\nu_{\rm u\,1}$ and $\nu_{\rm l\,1}$) in early phase,
while these QPOs disappear simultaneously in the late phase.  The
lower peak frequency $\nu_{\rm l\,1}$, which is the vertical
oscillation, corresponds to the Keplerian frequency $\nu_{\rm K\,1}$
at $r=6.3$ ({\it a dashed line}).  The upper peak frequency $\nu_{\rm
u\,1}$, which is the radial oscillation, corresponds to the sum of the
Keplerian frequency $\nu_{\rm K\,1}$ and the epicyclic frequency
$\nu_{\kappa\,1}$ ({\it a thin solid line}) at the same radius ({\it a
thin dotted line}).  The frequencies of the transient twin QPOs are
\begin{equation}
\nu_{\rm l\,1}=85~(10M_{\odot}/M)\hbox{[Hz]},
\end{equation}
\begin{equation}
\nu_{\rm u\,1}=152~(10M_{\odot}/M)\hbox{[Hz]}, 
\end{equation}
and the ratio of the peak frequencies is $\nu_{\rm l\,1}:\nu_{\rm
u\,1}=2:3.58$.

By comparing with (a) and (c), we find a pair of {\it persistent}
QPOs that are the radial oscillations (as indicated by the arrows
labeled as $\nu_{\rm u\,2}$ and $\nu_{\rm l\,2}$).  The lower peak
frequency $\nu_{\rm l\,2}$ corresponds to the Keplerian frequency
$\nu_{\rm K\,2}$ at $r=3.8$ ({\it a dashed line}).  The upper peak
frequency $\nu_{\rm u\,2}$ corresponds to the sum of the Keplerian
frequency $\nu_{\rm K\,2}$ and the epicyclic frequency
$\nu_{\kappa\,2}$ ({\it a thin solid line}) at the same radius ({\it a
thin dotted line}).  The frequencies of the persistent twin QPOs are
\begin{equation}
\nu_{\rm l\,2}=188~(10M_{\odot}/M)\hbox{[Hz]},
\label{plqpo:eqn}
\end{equation}
\begin{equation}
\nu_{\rm u\,2}=316~(10M_{\odot}/M)\hbox{[Hz]},
\label{puqpo:eqn}
\end{equation}
and the ratio of the peak frequencies is $\nu_{\rm l\,2}:\nu_{\rm
u\,2}=2:3.36$.

All the twin QPOs are excited near the resonant points as reported by
Kato (2004a).  We find the following empirical relations between the
upper frequency $\nu_{\rm u}$ and the lower frequency $\nu_{\rm l}$ in
these twin QPOs,
\begin{equation}
\nu_{\rm u}\approx \nu_{\rm K}+\nu_{\kappa},
\label{uqpo:eqn}
\end{equation}
\begin{equation}
\nu_{\rm l}\approx \nu_{\rm K},
\label{lqpo:eqn}
\end{equation}
where $\nu_{\rm K}$ and $\nu_{\kappa}$ is the Keplerian frequency and
the epicyclic frequency at the resonant radii, respectively.

\section{Discussion}

We performed three-dimensional MHD simulations of the radiatively
inefficient accretion flows around BHs.  We examined the time
variation of mass fluxes near the equatorial plane and found two twin
QPOs.  The peak frequency of the lower QPO corresponds to the
Keplerian frequency, whereas that of the upper QPO corresponds to the
sum of the Keplerian frequency and the epicyclic frequency at the same
radius. The ratio of these peak frequencies is close to 2:3.  In
addition, the flow structure is changed at the radii where the
oscillations are excited. Our results indicate that these twin QPOs
are most likely to be produced by the resonance between the Keplerian
frequency and the epicyclic frequency.

Kato (2004b) investigated that the excitation of the disk oscillations
by warps at the resonant radii $r\sim 3,\,3.63,\,4,\,6.46$ in the case
of Schwarzschild metric.  He reported that the resonant oscillation is
excited at the radius $r\sim 4$, on the other hand, it is dumped at
the radius $r\sim 6.46$.  He predicted that the twin QPOs is the
resonant oscillations at the radius $r=4$ where
$2\nu_{\kappa}=\nu_{\rm K}$ and their peak frequencies are equivalent
to $\nu_{\rm K}$ and $\nu_{\rm K}+\nu_{\kappa}=3\nu_{\kappa}$ at this
radius (see also Abramowicz et al. 2003; Rebusco 2004).

Our numerical study in the pseudo-Newtonian potential has a good
agreement with his results.  In fact, the flow structure is changed at
these resonant radii (see figure~\ref{fig1:eps}).  We have a similar
empirical relation of the peak frequencies of twin QPOs (see
equations~\ref{uqpo:eqn} and \ref{lqpo:eqn}) and the ratio of lower
and upper peak frequencies is close to 2:3.  In addition, the twin QPO
at $r=6.3$ is dumped in late phase, while that at $r=3.8$ is persisted
for the entire calculation time (see figure~\ref{fig3:eps}).

In the present calculation, the peak frequencies of twin QPOs are
slightly different from that in the resonant disk oscillation model.
The deviation of their peak frequency ratio from $2:3$ may be as a
result of the pseudo-Newtonian approximation since the maximum
epicyclic frequency is larger than that in the Schwarzschild metric by
about a factor of $\sqrt{2}$ (e.g., Okazaki, Kato, \& Fukue 1987).
For this reason, we expect that the peak frequency ratio is identical
to $2:3$ in general relativistic MHD calculations by using the
Schwarzschild metric.  Further numerical investigation of the detailed
internal disk structure of the MHD flows, such as fluctuations of
Maxwell stresses and magnetic reconnections would help clarify such
discrepancies.  Moreover, to elucidate the nature of several peaks in
the integrated PSD of our study [see figure~\ref{fig3:eps}~(c)],
detailed mode analysis in the magnetized disk is also important.  The
method described in this paper may provide a basis for investigating
the excitation mechanism of the disk oscillations in the MHD flows.

We can estimate masses of the non-rotating BHs by using the relation
of peak frequencies in the persistent twin QPOs [see
equations~(\ref{plqpo:eqn}) and (\ref{puqpo:eqn})] as
\begin{equation}
M_{\rm BH,l}=(\nu_{\rm l\,2}/\nu_{\rm l,obs})10M_{\odot}
\end{equation}
\begin{equation}
M_{\rm BH,u}=(\nu_{\rm u\,2}/\nu_{\rm u,obs})10M_{\odot}
\end{equation}
where $\nu_{\rm l,obs}$ ($\nu_{\rm u,obs}$) and $M_{\rm BH,l}$
($M_{\rm BH,u}$) is the observed lower (upper) peak frequency and the
mass of BHs estimated by using the lower (upper) QPO, respectively.
The estimated masses of BHs in two microquasars (see Remillard et
al. 2002) are shown in table~\ref{mass:tab}.  Since our calculation in
the pseudo-Newtonian potential overestimates both the epicyclic
frequency and the upper peak frequency, equation (17) indicates that
$M_{\rm BH,u}$ should be larger than $M_{\rm BH,l}$.  In addition, the
difference between $M_{\rm BH,l}$ and $M_{\rm BH,u}$ is important to
discriminate the spin of those BHs.  In the case of the rotating BHs,
we expect that the estimated masses of BHs should be larger than
$M_{\rm BH,l}$ and they may be converged at a unique value.

\begin{longtable}{rcccc}
  \caption{The estimated masses of BHs in two microquasars (Remillard
et al. 2002)}\label{mass:tab}
  \hline\hline
  Microquasars & $\nu_{\rm l,\,\hbox{obs}}$ [Hz] & $\nu_{\rm
u,\,\hbox{obs}}$ [Hz] & $M_{\rm BH,l}$ [$M_{\odot}$] & $M_{\rm BH,u}$
[$M_{\odot}$] \\
\endfirsthead
  \hline\hline
  Microquasars & $\nu_{\rm l,\,\hbox{obs}}$ [Hz] & $\nu_{\rm
u,\,\hbox{obs}}$ [Hz] & $M_{\rm BH,l}$ [$M_{\odot}$] & $M_{\rm BH,u}$
[$M_{\odot}$] \\
\endhead
  \hline
\endfoot
  \hline
\endlastfoot
  & & & & \\
  GRO J1655-40 & $300$ & $450$ & $6.26$ & $7.02$ \\
  XTE J1550-564 & $184$ & $276$ & $10.2$ & $11.4$ \\
\end{longtable}

Interestingly, the MHD flow structure is changed at these resonant
radii (see figure~\ref{fig1:eps}).  Hawley~\& Krolik (2001) also found
a similar flow pattern at $3\leq r\leq 15$ (see figure~9 in Hawley~\&
Krolik 2001).  Although they reported no evidence of QPOs in their PSD
of the mass accretion rate at ISCO, we could find the persistent twin
QPOs.  This is because our method has an advantage to evaluate where
QPOs are excited and what frequency they have simultaneously.

We also find that the radial oscillations are distributed in the
region between the radius of the maximum epicyclic frequency and that
of the inner boundary.  Although this indicates that the radial
oscillations are trapped in that region, we can find no strong
oscillations with the maximum epicyclic frequency as reported by the
previous studies (e.g., Matsumoto et al. 1988, 1989; Honma et
al. 1992).  

In comparison with the hydrodynamic study, we compute an effective
viscosity parameter $\alpha_{*}\equiv B_{r}B_{\phi}/4\pi p$ where
$B_{r}$ ($B_{\phi}$) and $p$ is the radial (toroidal) component of the
magnetic field and the gas pressure.  Since time-averaged
$\alpha_{*}\leq 0.1$ in that region, the viscous pulsational
instability may be marginally stable in the MHD flows (see Kato,
Honma,~\& Matsumoto 1988a, 1988b)

The power-law relations in the PSD of the present study (see
figure~\ref{fig3:eps}) is very similar to that of the aperiodic
X-ray fluctuations with $1/\nu$-like PSD (where $\nu$ is a frequency)
in some BHCs.  Takeuchi et al. (1995) constructed an accretion disk
model based on a cellular automaton model (e.g., Mineshige et
al. 1994) and successfully reproduced similar profiles of observed
PSD.  They asserted that the sporadic magnetic flares in the accretion
disks may be responsible to produce avalanche mass accretions in order
to satisfy their assumption of critical behavior.  Kawaguchi et
al. (2000) showed that the time fluctuations of the joule heating term
in the 3-D MHD simulation of Machida et al. (2000) produced a similar
power-law profile.  However, a direct connection between the magnetic
flares and the mass accretions has not been cleared yet.  Since the
variability of mass fluxes in our calculation may depends on the
magnetic stress induced locally in the disk, these may reflect the
spontaneous evolution of the MHD turbulence.  To elucidate their
connection, we need more careful studies.

The present paper assumes that the emissivity at various regions in
the disk is related to the gravitational energy dissipated by the
fluctuation of the mass fluxes.  Although the PSD in the present study
does not directly represent that of observed X-ray brightness
oscillations, our study confirms that a pair of frequencies can be
obtained by the resonance in the MHD disk.  The direct comparison with
observations is left as future work.

\vspace{5mm}

The author would like to thank S. Kato, R. Matsumoto, and S. Mineshige
for valuable discussions and K. Ohsuga, K. Watarai, and M. Takahashi
for stimulating discussions.  This work was supported in part by the
Grants-in Aid of the Ministry of Education, Science, Sports,
Technology, and Culture of Japan [15037202 (PI R. Matsumoto)].
Numerical computations were carried out on VPP5000 at the Astronomical
Data Analysis Center, ADAC, of the National Astronomical Observatory,
Japan (yyk27b, ryk22a).


\appendix


\begin{thebibliography}{}
\bibitem[Abramowicz and Kluzniak(2001)]{2001A&A...374L..19A}
Abramowicz, M.~A. and Klu{\' z}niak, W.\ 2001, \aap, 374, L19
\bibitem[Abramowicz et al.(2003)]{2003PASJ...55..467A} Abramowicz, M.~A., 
Karas, V., Kluzniak, W., Lee, W.~H., \& Rebusco, P.\ 2003, \pasj, 55,
467
\bibitem[Armitage \& Reynolds(2003)]{2003MNRAS.341.1041A} Armitage, 
P.~J.~\& Reynolds, C.~S.\ 2003, \mnras, 341, 1041
\bibitem[Balbus and Hawley(1991)]{1991ApJ...376..214B} Balbus,
S. A. \& Hawley, J. F. 1991 \apj, 376, 214
\bibitem[Chen \& Taam(1995)]{1995ApJ...441..354C} Chen, X.~\& Taam,
  R.~E.\ 1995, \apj, 441, 354
\bibitem[Ford et al.(1997)]{1997ApJ...475L.123F} Ford, E., Kaaret, P.,
Tavani, M., Barret, D., Bloser, P., Grindlay, J., Harmon, B.~A.,
Paciesas, W.~S., and Zhang, S.~N.\ 1997, \apjl, 475, 123
\bibitem[Fukue and Okada(1990)]{1990PASJ...42..533F} Fukue, J and
Okada, R.\ 1990, \pasj, 42, 533
\bibitem[Hasinger(1987)]{1987A&A...186..153H} Hasinger, G.\ 1987,
\aap, 186, 153
\bibitem[Hawley(2000)]{2000ApJ...528..462H} Hawley, J.~F.\ 2000, \apj,
528, 462
\bibitem[Hawley \& Krolik(2001)]{2001ApJ...548..348H} Hawley, J.~F.~\& 
Krolik, J.~H.\ 2001, \apj, 548, 348
\bibitem[Hawley, Balbus, and Stone(2001)]{2001ApJ...554L..49H} Hawley,
J.~F., Balbus, S.~A., and Stone, J.~M.\ 2001, \apjl, 554, 49
\bibitem[Hawley and Balbus(2002)]{2002ApJ...573..738H} Hawley,
J.~F. and Balbus, S.~A.\ 2002, \apj, 573, 738
\bibitem[Hirose and Osaki(1990)]{1990PASJ...42..135H} Hirose, M. and
Osaki, Y.\ 1990, \pasj, 42, 135
\bibitem[Hirose, Osaki, Mineshige(1991)]{1991PASJ...43..809H} Hirose,
M., Osaki, Y., and Mineshige, S.\ 1991, \pasj, 43, 809
\bibitem[Honma, Matsumoto, \& Kato(1992)]{1992PASJ...44..529H} Honma, F., 
Matsumoto, R., \& Kato, S.\ 1992, \pasj, 44, 529
\bibitem[Igumenshchev, Narayan, and
Abramowicz(2003)]{2003ApJ...592.1042I} Igumenshchev, I.~V., Narayan,
R., and Abramowicz, M.~A.\ 2003, \apj, 592, 1042
\bibitem[Kato(2001)]{2001PASJ...53....1K} Kato, S.\ 2001, \pasj, 53, 1
\bibitem[Kato(2003)]{2003PASJ...55..801K} Kato, S.\ 2003, \pasj, 55,
801
\bibitem[Kato(2004a)]{2004PASJ...56..559K} Kato, S.\ 2004a, \pasj, 56, 559
\bibitem[Kato(2004b)]{} Kato, S.\ 2004b, \pasj (submitted)
\bibitem[Kato, Honma, \& Matsumoto(1988a)]{1988MNRAS.231..37K} Kato,
S. , Honma, F., \& Matsumoto, R.\ 1988a, \mnras, 231, 37
\bibitem[Kato, Honma, \& Matsumoto(1988b)]{1988PASJ...40..709K} Kato,
S. , Honma, F., \& Matsumoto, R.\ 1988b, \pasj, 40, 709
\bibitem[Kato, Fukue, \& Mineshige(1998)]{1998bhad.conf.....K} 
Kato, S., Fukue, J., \& Mineshige, S.\ 1998, Black-hole accretion
disks.~Edited by Shoji Kato, Jun Fukue, and Sin Mineshige.~ Publisher:
Kyoto, Japan: Kyoto University Press, 1998.~ISBN: 4876980535,
Chap. 13, 14, and 15
\bibitem[Kato, Mineshige, and Shibata(2004)]{2004ApJ...605..307K}
Kato, Y., Mineshige, S., and Shibata, K.\ 2004, \apj, 605, 307 (KMS04)
\bibitem[Kawaguchi et al(2000)]{2000PASJ...52L...1K} Kawaguchi, T.,
Mineshige, S., Machida, M., Matsumoto, R., and  Shibata, K.\ 2000,
\pasj, 52, L1
\bibitem[Lubow(1991)]{1991ApJ...381..259L} Lubow, S.~H.\ 1991, \apj, 381, 
259
\bibitem[Lubow(1991)]{1991ApJ...381..268L} Lubow, S.~H.\ 1991, \apj, 381, 
268
\bibitem[Machida, Hayashi, and Matsumoto(200)]{2000ApJ...532L..67M}
Machida, M., Hayashi, M.~R., and Matsumoto, R.\ 2000, \apjl, 2000,
532, 67
\bibitem[Machida and Matsumoto(2003)]{2003ApJ...585..429M} Machida,
M. and Matsumoto, R.\ 2003, \apj, 585, 429
\bibitem[Matsumoto, Kato, and Honma(1989)]{1989tad..conf..167M}
Matsumoto, R., Kato, S., and Honma, F.\ 1989, NATO ASIC Proc. 290:
Theory of Accretion Disks, 167
\bibitem[Matsumoto, Kato, \& Honma(1988)]{1988pnsb.conf..155M} Matsumoto, 
R., Kato, S., \& Honma, F.\ 1988, Physics of Neutron Stars and Black Holes, 
155
\bibitem[Mendez et al.(1998)]{1998ApJ...505L..23M} Mendez, M., van der 
Klis, M., Wijnands, R., Ford, E.~C., van Paradijis, J., \& Vaughan,
B.~A.\ 1998, \apjl, 505, L23
\bibitem[Menou(2003)]{2003ApJ...596..414M} Menou, K.\ 2003, \apj, 596,
414
\bibitem[Milsom \& Taam(1996)]{1996MNRAS.283..919M} Milsom, J.~A.~\&
  Taam, R.~E.\ 1996, \mnras, 283, 919
\bibitem[Milsom \& Taam(1997)]{1997MNRAS.286..358M} Milsom, J.~A.~\&
  Taam, R.~E.\ 1997, \mnras, 286, 358
\bibitem[Mineshige, Takeuchi, and
Nishimori(1994)]{1994ApJ...435L.125M} Mineshige, S., Takeuchi, M., and
Nishimori, H.\ 1994, \apjl, 435, 125
\bibitem[Morgan, Remillard, and Greiner(1997)]{1997ApJ...482..993M}
Morgan, E.~H., Remillard, R.~A., and Greiner, J.\ 1997, \apj, 482, 993
\bibitem[Nowak and Wagoner(1991)]{1991ApJ...378..656N} Nowak,
M.~A. and Wagoner, R.~V.\ 1991, \apj, 378, 656
\bibitem[Nowak and Wagoner(1992)]{1992ApJ...393..697N} Nowak,
M.~A. and Wagoner, R.~V.\ 1992, \apj, 393, 697
\bibitem[Ogilvie(2003)]{2003MNRAS.340..969O} Ogilvie, G.~I.\ 2003, \mnras, 
340, 969
\bibitem[Okazaki, Kato, Fukue(1987)]{1987PASJ...39..457O} Okazaki,
A.~T., Kato, S., and Fukue, J.\ \pasj, 1987, 39, 457
\bibitem[Paczynski and Wiita(1980)]{1980A&A....88...23P}
Paczy\'{n}sky, B. and Wiita, P.~J.\ 1980, \aap, 1980, 88, 23
\bibitem[Rebusco(2004)]{} Rebusco, P.\ 2004, \pasj (in print)
\bibitem[Remillard et al.(1999b)]{1999ApJ...522..397R} Remillard,
R.~A., Morgan, E.~H., McClintock, J.~E., Bailyn, C.~D., and Orosz,
J.~A.\ 1999, \apj, 522, 397
\bibitem[Remillard et al.(1999a)]{1999ApJ...517L.127R} Remillard,
R.~A., McClintock, J.~E., Sobczak, G.~J., Bailyn, C.~D., Orosz, J.~A.,
Morgan, E.~H., and Levine, A.~M.\ 1999, \apjl, 517, 127
\bibitem[Remillard et al.(2002)]{2002ApJ...580.1030R} Remillard,
R.~A., Muno, M.~P., McClintock, J.~E., and Orosz, J.~A.\ 2002, \apj,
580, 1030
\bibitem[Sano, Inutsuka, Turner, \& Stone(2004)]{2004ApJ...605..321S} Sano, 
T., Inutsuka, S., Turner, N.~J., \& Stone, J.~M.\ 2004, \apj, 605, 321
\bibitem[Shakura and Sunyaev(1973)]{1973A&A....24..337S} Shakura,
N.~I.~\& Sunyaev, R.~A.\ 1973, \aap, 24, 337
\bibitem[Strohmayer, Zhang, Swank, Smale, Titarchuk, Day, and
Lee(1996)]{1996ApJ...469L...9S} Strohmayer, T.~E., Zhang, W., Swank,
J.~H., Smale, A., Titarchuk, L., Day, C., and Lee, U.\ 1996, \apjl,
468, 9
\bibitem[Takeuchi, Mineshige, Negoro(1995)]{1995PASJ...47..617T}
Takeuchi, M., Mineshige, S., and Negoro, H.\ 1995, \pasj, 47, 617
\bibitem[van der Klis(1999)]{1999ptgr.conf..259V} van der Klis, M.\ 1999, 
Pulsar Timing, General Relativity and the Internal Structure of Neutron 
Stars, 259
\bibitem[van der Klis(2000)]{2000ARA&A..38..717V} van der Klis, M.\
2000, \araa, 38, 717
\bibitem[van der Klis, Wijnands, Horne, and
Chen(1997)]{1997ApJ...481L..97V} van der Klis, M., Wijnands, R.~A.~D.,
Horne, K., and Chen, W.\ 1997, \apjl, 481, 97
\bibitem[Whitehurst(1988)]{1988MNRAS.232...35W} Whitehurst, R.\ 1988,
\mnras, 232, 35
\bibitem[Wijnands et al.(1997)]{1997ApJ...490L.157W} Wijnands, R.,
Homan, J., van der Klis, M., Mendez, M., Kuulkers, E., van Paradijs,
J., Lewin, W.~H.~G., Lamb, F.~K., Psaltis, D., and Vaughan, B.\ 1997,
\apjl, 490, 157
\bibitem[Zhang, Smale, Strohmayer, \& Swank(1998)]{1998ApJ...500L.171Z} 
Zhang, W., Smale, A.~P., Strohmayer, T.~E., \& Swank, J.~H.\ 1998, \apjl, 
500, L171
\end{thebibliography}
\end{document}